\documentclass{article}
\usepackage{spconf,amsmath}
\usepackage{amsthm}
\usepackage{amsfonts}
\usepackage{amssymb}
\usepackage{mathrsfs}
\usepackage{mathtools}
\usepackage[english]{babel}
\usepackage{float}
\usepackage[toc,page]{appendix}
\usepackage[pdftex]{graphicx}
\usepackage{epstopdf}
\usepackage{amsmath}
\usepackage{color}
\usepackage{multirow}
\usepackage{graphicx}
\usepackage{IEEEtrantools}
\usepackage{bm}
\usepackage{balance}
\usepackage{microtype}
\usepackage[caption=false,font=footnotesize]{subfig}
\usepackage{amsthm, thmtools}
\usepackage{mathtools}
\usepackage{hyperref}
\newtheorem{theorem}{Theorem}
\usepackage[most]{tcolorbox}

\makeatletter
\def\smalloverbrace#1{\mathop{\vbox{\m@th\ialign{##\crcr\noalign{\kern3\p@}%
  \tiny\downbracefill\crcr\noalign{\kern3\p@\nointerlineskip}%
  $\hfil\displaystyle{#1}\hfil$\crcr}}}\limits}
\makeatother
\makeatletter
\def\smallunderbrace#1{\mathop {\vtop {\m@th \ialign {##\crcr $\hfil \displaystyle {#1}\hfil $\crcr \noalign {\kern 3\p@ \nointerlineskip } \tiny\upbracefill \crcr \noalign {\kern 3\p@ }}}}\limits}
\makeatother

\DeclareMathOperator*{\argmin}{arg\,min}
\usepackage{cite}
\usepackage{comment}
\allowdisplaybreaks

\numberwithin{equation}{section}
\usepackage{bbm}
\usepackage{fancyvrb}
\makeatletter
\newcommand{\settheoremtag}[1]{
  \let\oldthetheorem\thetheorem
  \renewcommand{\thetheorem}{#1}
  \g@addto@macro\endtheorem{
    \addtocounter{theorem}{-1}
    \global\let\thetheorem\oldthetheorem}
  }
\makeatother

\usepackage[switch]{lineno}
\usepackage[named]{algo}
\usepackage[noend]{algpseudocode}
\usepackage{algorithm}

\DeclareMathOperator*{\minimize}{minimize }

\ninept

\title{Learning to reconstruct signals with  Inexact sensing operator via knowledge distillation} 
	
\name{{Roman Jacome}$^\dagger$, {Leon Suarez}$^\ddagger$, {Romario Gualdr\'on-Hurtado}$^\ddagger$, Luis Gonzalez$^\ddagger$, Henry Arguello$^\ddagger$}
\address{$^\dagger$Department of Electrical, Electronics and Telecommunications Engineering \\
$^\ddagger$Department of Systems and Informatics Engineering \\Universidad Industrial de Santander, Bucaramanga, Colombia, 680002\\\thanks{This work was supported by project VIE-UIS under the research project 3944 and ICETEX and MINCIENCIAS through the CTO 2022-0716 Sistema óptico-computacional tipo pushbroom en el rango visible e infrarrojo cercano (VNIR), para la clasificación de frutos cítricos sobre bandas transportadoras mediante aprendizaje profundo, desarrollado en alianza con citricultores de Santander, under Grant 8284. Romario Gualdrón acknowledges the support of the IEEE SPS Scholarship.}}

\begin{document}
\ninept
\setlength{\abovedisplayskip}{3pt}
\setlength{\belowdisplayskip}{3pt}
\maketitle
\vspace{-8pt}
\begin{abstract}

In computational optical imaging and wireless communications, signals are acquired through linear coded and noisy projections, which are recovered through computational algorithms. Deep model-based approaches, i.e., neural networks incorporating the sensing operators, are the state-of-the-art for signal recovery. However, these methods require exact knowledge of the sensing operator, which is often unavailable in practice, leading to performance degradation. Consequently, we propose a new recovery paradigm based on knowledge distillation. A teacher model, trained with full or almost exact knowledge of a synthetic sensing operator, guides a student model with an inexact real sensing operator. The teacher is interpreted as a relaxation of the student since it solves a problem with fewer constraints, which can guide the student to achieve higher performance.  We demonstrate the improvement of signal reconstruction in computational optical imaging for single-pixel imaging with miscalibrated coded apertures systems and multiple-input multiple-output symbols detection with inexact channel matrix.
\end{abstract}
\vspace{-0.2cm}
\vspace{-0.35cm}
\section{Introduction}\vspace{-0.25cm}
In several applications, signals $\mathbf{x}\in\mathbb{R}^{n}$ are acquired through linear coded {and noisy} projections $\mathbf{y} \in \mathbb{R}^m$ as {following}
    $\mathbf{y} = \mathbf{A}\mathbf{x} + \boldsymbol{\omega}$, where $\mathbf{A}\in\mathbb{R}^{m\times n}$ is the sensing operator and $\boldsymbol{\omega} \in \mathbb{R}^m$ is additive Gaussian noise. {Consequently}, a reconstruction algorithm is required to retrieve the underlying signal after the acquisition process. This computational sensing framework has opened new frontiers in a wide range of wireless communications {\cite{DL_COMMUNICATIONS,he2020model,he2019model}} and imaging applications, including {microscopy \cite{mcleod2016unconventional, PSF_ESTIMATION_SHIFT_VARIANT}, computational photography \cite{lukac2017computational}, {and} remote sensing \cite{erkmen2012computational}}.  Therefore, recovering the underlying signal {is} a long-standing and widely studied problem,  typically referred to as an (ill-posed) inverse problem.  Several methods are based on model-based approaches, algorithms that take into account the sensing model. These methods are formulated as the optimization of a data-fidelity term $f(\mathbf{x})$ aiming to fit the estimated signal to the observed measurements, and a regularizer term {$g(\mathbf{x})$} to promote prior information {about} the signal. The optimization problem is then formulated as
\begin{equation}
    \minimize_{\tilde{\mathbf{x}}} f(\tilde{\mathbf{x}})+\lambda g(\mathbf{\tilde{x}}),\label{eq:opt_org}
\end{equation}
where $\lambda$ is the regularization parameter  {that} balances the fidelity and regularization terms. For the regularization term $g(\mathbf{x})$, a plethora of approaches {has} been {proposed such as sparsity, total variation or low-rank \cite{jin2017sparsity,osher2005iterative,dong2014compressive}}. While these methods rely on hand-crafted designs of the signal prior, in some practical scenarios, these assumptions {are not met leading to subpar} performance \cite{yang2018admm}. Learning-based approaches have also been proposed, where the signal prior is implicitly optimized using a large paired signal-measurement dataset \cite{gualdron2024deep}. Mainly, these methods aim to fit a neural network that maps from the {measurements} to the {target signal.} 
The network structure is one of the most important aspects of this kind of recovery method. Commonly used architectures such as convolutional or fully connected neural networks, and transformers,  are considered \textit{black boxes} due to their lack of {interpretability \cite{monga2021algorithm}}. 
Unrolling networks, {a model-based} neural network, address this issue by providing a hybrid solution that integrates the structured iterative algorithms of model-driven approaches with the adaptability of deep learning, which not only enhances interpretability but also delivers high performance in recovery tasks {\cite{zhao2022model,shlezinger2023model,monga2021algorithm}}. These model-based methods {require} a {complete} knowledge of the sensing operator $\mathbf{A}$, which in some applications may not be available. {In this work,} we consider two main applications where this problem is crucial:

\noindent \;\textit{Computational optical imaging:} In a single-pixel camera (SPC), a coded aperture (CA) (a binary array with values $\{0,1\}$ or $\{-1,1\}$) modulates the scene spatially. This has applications in various imaging modalities such as spectral, depth, and x-ray imaging \cite{gibson2020single}. However, the sensing matrix is always considered ideal i.e., binary. However, in practice, the CA is implemented in devices with some non-idealities that attenuate the light such as digital micromirror devices \cite{garcia2023calibration}. In other computational optical imaging applications appears this problem such as coded snapshot spectral imaging with coded apertures or with diffractive optical elements \cite{gualdron2024deep,li2022quantization}.

  \noindent \;\textit{Wireless communications:} In multiple-input multiple-output (MIMO) systems we want to recover symbols that are transmitted from linear memoryless channels {\cite{MIMO_SURVEY,solano2024data,samuel2017deep,he2020model}}. {Before} the symbol detection process, a channel estimation method is performed using some transmitted pilot (known) symbols \cite{he2020model}. Usually, the channel estimation contains errors {with respect to} the ground truth channel leading to a mismatch in the symbol detection algorithm \cite{he2020model}. This phenomenon appears in dynamic communications channels in integrated sensing and communications applications \cite{vargas2023dual,jacome2024multi}. 
  
To address these issues, several works have focused on estimating the unknown sensing operator {\cite{song2022high,gualdron2024deep,campo2024point}}. This traces back to blind-deconvolution or {blind-identification problems \cite{ren2020neural,vargas2023dual,ahmed2013blind}}. However, jointly estimating the sensing operator and the underlying signal is a highly challenging problem. This work introduces a new paradigm to address the sensing operator mismatch in deep model-based recovery methods. The proposed method is inspired {by} the widely used knowledge distillation (KD) technique in deep neural networks \cite{hinton2015distilling,gou2021knowledge}.  KD was originally introduced to train small and {resource-constrained} models (student) with the guidance of a, usually pre-trained, big and {high-performance} model (teacher). The idea is that the teacher can transfer his knowledge to the student to improve his performance (for more details on KD see \cite{gou2021knowledge}). 

In this work, we extrapolate the KD concept using a model with a fully or almost fully characterized sensing operator (the teacher model). {This sensing operator is defined only in simulation, since it may not model the real-world acquisition phenomenon but allows a high reconstruction performance. The student model uses a less characterized sensing operator, i.e., with a partial knowledge or inexact sensing operator, which follows a realistic sensing scenario.} The teacher model system serves as a guide for the student model, thereby improving its performance. To transfer the knowledge from the teacher model to the student model we proposed two distillation loss functions, one that matches the gradient steps of the unrolling network at each stage, and the second promotes that the student has similar signal estimation at each stage of the network. Here, different from the traditional KD application in neural networks, the teacher and student models are of the same size, i.e., they have the same number of parameters, the only difference is the sensing operator available in the model. {The student model, while matching the teacher in parameters and layers, is limited by its incomplete knowledge of the sensing operator, affecting its performance. This paper reinterprets the KD concept to address this specific problem.}
We validate this approach in two important applications: computational optical imaging for SPC reconstruction and MIMO detection in wireless communications. For the first one, we consider a teacher model trained with all the ideal or almost-ideal SPC sensing matrix (binary numbers) with the unrolling network described in \cite{suarez2024highly} and the student only has access to the miscalibrated SPC sensing matrix, which is the binary matrix plus additive noise. In the MIMO detection problem, we employed the DetNet architecture \cite{samuel2017deep}, where the teacher is the trained DetNet with near-perfect knowledge of the channel state information (CSI) while the student has an inexact knowledge of the CSI.
\label{sec:intro}\vspace{-0.5cm}
\section{Deep {Model-Based} Recovery Background}\vspace{-0.3cm}
{Given the} success of deep model-based neural networks {in recovery tasks}, { such as}
unrolling networks,  in this work, we use this {approach} as the backbone {of our} recovery methods. In general, we want to learn a neural network that uses the sensing operator to map the measurements to the underlying signal {$\mathcal{M}_{\boldsymbol{\theta}}(\cdot):\mathbb{R}^{m} \times \mathbb{R}^{m\times n} \rightarrow \mathbb{R}^{n}$ with parameters $\boldsymbol{\theta}$. The network is trained as
\begin{equation}
    \theta^* = \argmin_{\theta} \mathbb{E}_{\mathbf{x},\mathbf{y}} \left[ \ell(\mathcal{M}_\theta(\mathbf{y},\mathbf{A}),\mathbf{x})\right],\label{eq:network_opt}\vspace{-0.1cm}
\end{equation}
where $\ell(\cdot,\cdot)$ is the loss function. This problem is solved using off-the-shelf stochastic gradient descent algorithms \cite{monga2021algorithm}. Unrolling networks are built as sequential models of $L$ layers, where each layer is interpreted as an iteration of an iterative  algorithm as:
\begin{equation}
    \mathcal{M}_\theta({\mathbf{y},\mathbf{A}}) = \mathcal{M}_{\theta^L}(\mathcal{M}_{\theta^{L-1}}(\cdots \mathcal{M}_{\theta^1}(\mathbf{y},\mathbf{A}),\mathbf{A}),\mathbf{A}).\label{eq:structure}
\end{equation}
Mainly, we use two different unrolling networks, an alternating direction method of multipliers (ADMM) \cite{boyd2011distributed} based network used for the SPC experiments. The second is the {DetNet \cite{li2018detnet}} which is {an unrolling} network inspired {by} a projected gradient descent method, this network is used for the MIMO detection in the wireless communications application. \vspace{-0.45cm}
\subsection{{ADMM-Inspired} Network for SPC recovery}\vspace{-0.2cm}
First, let's define the sensing process for SPC. Here $\mathbf{x}$ is the vectorized image, we only consider grayscale images but it can be extended for color or spectral images \cite{gibson2020single}. The rows of the sensing operator are the vectorized coded apertures $\mathbf{a}_i \in \{-1,1\}^n, i=1,\dots,m$ leading to $\mathbf{A} = \left[ \mathbf{a}_1, \dots, \mathbf{a}_n\right]^\top$, where $m$ is the number of snapshots. The ADMM formulation {divides the problem \eqref{eq:opt_org} into small steps}. Thus, an auxiliary variable $\mathbf{z} \in \mathbb{R}^n$ is introduced, leading to
\begin{equation}
    \minimize_{\tilde{\mathbf{x}},\mathbf{z}} \frac{1}{2}\Vert\mathbf{y-A\tilde{x}}\Vert_2^2 + \lambda g({\mathbf{z}}) \;\text{ s.t. } \mathbf{z} = \mathbf{\tilde{x}}. \label{eq: opt_2}
\end{equation}

To solve this problem for both variables, $\mathbf{x}$ and $\mathbf{z}$, the augmented Lagrangian is formulated as:
\begin{equation}
    L_\rho(\tilde{\mathbf{x}},\mathbf{z},\mathbf{u}) = \frac{1}{2}\Vert\mathbf{y-A\tilde{x}}\Vert_2^2 + \lambda g({\mathbf{z}})+ \frac{\rho}{2}\Vert \mathbf{z} - \mathbf{\tilde{x}} + \mathbf{u}\Vert,
\end{equation}
where $\rho>0$ is a penalty parameter, and $\mathbf{u} \in \mathbb{R}^n$ is the Lagrange multiplier. Then, the optimization problem is solved iteratively by updating the variables $\mathbf{z}$, $\tilde{\mathbf{x}}$, and $\mathbf{u}$, following
\begin{equation}
\begin{cases}
    \mathbf{z}^{l+1} &:=  \arg \min _{\mathbf{z}} \left( g(\mathbf{z}) + \frac{\rho}{2} \left\| \mathbf{\tilde{x}}^l - \mathbf{z} + \mathbf{u}^l \right\|_2^2 \right), \\
     \mathbf{\tilde{x}}^{l+1} &:= \arg \min _{\mathbf{\tilde{x}}} \left( f(\mathbf{\tilde{x}}) + \frac{\rho}{2} \left\| \mathbf{\tilde{x}} - \mathbf{z}^{l+1} + \mathbf{u}^l \right\|_2^2 \right), \\
    \mathbf{u}^{l+1} &:= \mathbf{u}^l + \mathbf{\tilde{x}}^{l+1} - \mathbf{z}^{l+1}.
\end{cases}
\label{eq:iterative_way}
\end{equation}
To solve the {$\mathbf{z}$-update} a neural network $\mathcal{D}_{\beta^l} (\cdot)$ is employed to {learn} the proximal operator of this term. The {$\tilde{\mathbf{x}}$-update} is performed via gradient descent formulation. And finally, the dual variable $\mathbf{u}$ is updated via dual ascent iteration.  {In Algorithm \ref{alg:admm}, the} unfolding of the ADMM iterations. Highlighted in \textcolor{blue}{blue} are the learnable parameters at every iteration which are $\theta = \{\beta^l, \alpha^l, \rho^l\}_{l=1}^{L}$. \vspace{-0.45cm}
\subsection{DetNet for MIMO detection}\vspace{-0.15cm}
MIMO detection problem refers to recovering the signal $\overline{\mathbf{x}}\in \mathbb{C}^{\overline{n}}$ using the received signal $\overline{\mathbf{y}}\in\mathbb{C}^{\overline{m}}$ and the known channel matrix $\mathbf{A}$. 
 Let $\overline{\mathbf{A}} \in \mathbb{C}^{\overline{m}
    \times \overline{n}}$
    is the channel matrix with 
    $\overline{m} \leq \overline{n}$ where $\overline{n}$ and $\overline{m}$ are the number of transmitter and receiver antennas. We consider the MIMO system model    $\mathbf{\overline{y} = \overline{A}\overline{x} + \overline{\boldsymbol{w}}}$,    where $\mathbf{x} \in \mathcal{S}^{\overline{n}}$ is a transmitted signal from finite square signal constellation $\mathcal{S}$, and $\overline{\mathbf{w}} \in \mathbb{C}^{\overline{m}}$ is a Gaussian noise vector. The complex-valued model can be reformulated as the real-valued model $   \mathbf{y = Ax + \boldsymbol{\omega}},$    where
    \begin{align} & \mathbf{y}=\left[\begin{smallmatrix}\mathfrak{R}(\overline{\mathbf{y}}) \\ \mathfrak{I}(\overline{\mathbf{y}})\end{smallmatrix}\right],  \mathbf{x}=\left[\begin{smallmatrix}\mathfrak{R}(\overline{\mathbf{x}}) \\ \mathfrak{I}(\overline{\mathbf{x}})\end{smallmatrix}\right],   \boldsymbol{\omega}=\left[\begin{smallmatrix}\mathfrak{R}(\overline{ \boldsymbol{\omega}}) \\ \mathfrak{I}(\overline{\mathbf{w}})\end{smallmatrix}\right] \mathbf{A}=\left[\begin{smallmatrix}\mathfrak{R}(\overline{\mathbf{A}}) & -\mathfrak{I}(\overline{\mathbf{A}}) \\ \mathfrak{I}(\overline{\mathbf{A}}) & \mathfrak{R}(\overline{\mathbf{A}})\end{smallmatrix}\right],\nonumber\end{align}
    with $m = 2\overline{m}$ and $n = 2\overline{n}$, therefore the receiver can be rearranged to obtain the equivalent real channel model (2).
    Where $\mathbf{y}\in\mathbb{R}^{m}$, $\mathbf{A} \in \mathbb{R}^{m\times n}$, $\mathbf{x} \in \mathcal{S}^{n}$ and $\mathbf{w} \in \mathbb{R}^{m}$ are the received signal, the channel matrix, the transmitted signal and the noise, respectively. In this case, $\mathcal{S} = \mathfrak{R}(\overline{\mathcal{S}})$ represent the real part of complex constellations.
In the MIMO detection problem, since the underlying signal (symbols) belongs to a defined constellation (BPSK, {QPSK, or} 16QAM among others), i.e., $\mathbf{x}\in\mathcal{S}^{n}$, where in the case of BPSK $\mathcal{S} = \{-1,1\}$.  Thus, the optimization problem is formulated as:
\begin{equation}
    \minimize_{\tilde{\mathbf{x}}} \Vert \mathbf{y-A\tilde{x}}\Vert_2 + I_{\mathcal{S}}(\mathbf{\tilde{x}}),\label{eq:mimo_opt}
\end{equation}
where $I_{\mathcal{S}}(\cdot)$ is an element-wise indicator function over the constellation set $\mathcal{S}$, i.e., $I_{\mathcal{S}}(x) = 0 $ if $\mathbf{x}\in\mathcal{S}$ and $I_{\mathcal{S}}(x) = \infty$ if $x\not\in\mathcal{S}$.  Thus, a projected gradient descent (PGD) approach is highly suitable for this problem. Mainly, PGD iteration for \eqref{eq:mimo_opt} is 
\begin{equation}
    \mathbf{\tilde{x}}^{l+1} = \Pi_{I_\mathcal{S}}\left(\tilde{\mathbf{x}}^l -\alpha^l \mathbf{A}^\top\left(\mathbf{A\tilde{x}}^l-\mathbf{y}\right)\right).
\end{equation}
The DetNet improves this algorithm by lifting the gradient step to a higher dimension using fully connected layers and non-linear activation functions. This is followed by two fully connected layers. The summary of the algorithm is shown in Algorithm \ref{alg:detnet}. The function $\varrho(\cdot)$ is the ReLu activation function and the $\psi_t(\cdot) = -1 + \frac{\varrho(\cdot + t)}{\vert t \vert} - \frac{\varrho(\cdot - t)}{\vert t \vert} $ is the piecewise linear soft sign operator that works as the proximal operator over the constellation set. Here the trainable parameters are $\theta =\{\mathbf{W}_{1}^l,\mathbf{b}_{1}^l,\mathbf{W}_{2}^l,\mathbf{b}_2^l,\mathbf{W}_3^l,\mathbf{b}_3^l\}_{l=1}^L$.\vspace{-0.3cm}
\setlength{\textfloatsep}{0pt}
\begin{algorithm}[!t]
\caption{{ADMM-Inspired} Network}\label{alg:cap}
\begin{algorithmic}[1]
\Require $\mathbf{A}$, $\mathbf{y}$, $L$
\State $\mathbf{x}^0 = \mathbf{A}^T\mathbf{y},   \mathbf{u}^0 = \mathbf{0}$, \Comment{Network input initialization}
\For {$l=1:L$}
  \State  $\mathbf{z}^{l+1}:=\mathcal{D}_{\textcolor{blue}{\beta^{l+1}}}\left(\mathbf{\tilde{x}}^{l}+\mathbf{u}^l\right).$ \Comment{Auxiliary variable update}

        \State $\mathbf{\tilde{x}}^{l+1} := \mathbf{\tilde{x}}^l - \textcolor{blue}{\alpha^{l+1}} (\overbrace{\mathbf{A}^\top \left( \mathbf{A} \mathbf{\tilde{x}}^l - \mathbf{y} \right)}^{=\nabla f(\tilde{\mathbf{x}}^{l})}   \quad $ \Comment{$\tilde{\mathbf{x}}$-Update}
        
        $\quad\quad\quad- \textcolor{blue}{\rho^{k+1}} \left( \mathbf{\tilde{x}}^l - \mathbf{z}^{l+1} + \mathbf{u}^l \right) \Big)$
    \State $\mathbf{u}^{l+1} = \mathbf{u}^{l}+\mathbf{x}^{l+1}-\mathbf{z}^{l+1}$ \Comment{${\mathbf{u}}$-Update}
\EndFor
\end{algorithmic}
\label{alg:admm}
\end{algorithm}
\setlength{\textfloatsep}{2pt}
\begin{algorithm}[!t]
\caption{MIMO detection network}
\begin{algorithmic}[1]
\Require $\mathbf{A}$, $\mathbf{y}$, $L$
\State $\mathbf{x}^0 = \mathbf{0}, \mathbf{v}^0=   \mathbf{0}$, \Comment{Network input initialization}
\For {$l=1:L$}
    
  \State  $\mathbf{z}^l = \varrho\left(\overbrace{\textcolor{blue}{\mathbf{W}_{1}^l}\left[\begin{smallmatrix}
      \mathbf{A}^\top\mathbf{y}\\
      \tilde{\mathbf{x}}^l\\
      \mathbf{A}^\top\mathbf{A\tilde{x}}^l\\
      \mathbf{v}^l
  \end{smallmatrix}\right]}^{=\nabla f(\tilde{\mathbf{x}}^{l})} + \textcolor{blue}{\mathbf{b}_{1}^l}\right)$ \Comment{Gradient descent}
  \State $\mathbf{\tilde{x}}^{l+1} = \psi_t(\textcolor{blue}{\mathbf{W}_{2}^l}\mathbf{z}^l + \textcolor{blue}{\mathbf{b}_{2}^l})$ \Comment{Soft-sign operator}

  \State $\mathbf{v}^{l+1} = \textcolor{blue}{\mathbf{W}_{3}^l}\mathbf{z}^l + \textcolor{blue}{\mathbf{b}_{3}^l}$ \Comment{Lifted variable Update}
\EndFor
\end{algorithmic}
\label{alg:detnet}
\end{algorithm}
\vspace{0.2cm}\section{Signal Reconstruction with Inexact Sensing Operator}\vspace{-0.3cm}
Unrolling methods, as described in the previous section, require full knowledge of the sensing operator $\mathbf{A}$ to have optimal reconstruction performance. Here, we consider that the sensing operator is composed as  $\mathbf{A} = \mathbf{A}_k + \mathbf{A}_u$, where $\mathbf{A}_k$ is the known portion of the sensing operator and $\mathbf{A}_u$ is the unknown part. In unrolling networks, having only access to an inexact sensing operator affects the fidelity term update  since we only have that
\begin{equation}
    f(\tilde{\mathbf{x}}) = \Vert\smalloverbrace{\mathbf{y}}^{\tiny{\clap{$\mathbf{(A_k+A_u)x}+\boldsymbol{\omega}$}}} - \mathbf{A}_k\mathbf{\tilde{x}}\Vert,\label{eq: mismatched}
\end{equation}
this mismatch significantly affects the gradient computation of the fidelity term, {specifically on} line 4 in Algorithm \ref{alg:admm} and line 5 in Algorithm \ref{alg:detnet}, since the direction of the gradient {leads to the wrong} optimization direction. Additionally, if the power of the unknown matrix $\mathbf{A}_u$ is significantly higher than the power of the known sensing matrix the performance drop is higher.  Here we consider that the unknown sensing operator is a random variable, where each entry is i.i.d from a Gaussian distribution $[\mathbf{A}_u]_{i,j} \sim \mathcal{N}(0,\sigma^2)$. Here, {the power of the unknown sensing operator is directly proportional to the variance $\sigma^2$.} As a proof-of-concept of this issue, we validate on both applications (MIMO detection and SPC recovery) the effect of the power of the unknown sensing matrix in the models' performance. First, for the MIMO detection, we run the DetNet model (Algorithm \ref{alg:detnet}) for different variances of the unknown signal and plot the bit error rate (BER, the lower the better) of the recovered BPSK signal and for different signal-to-noise-ratio (SNR) in the measurements. The results shown in Fig. \ref{fig:sigma}a) confirm the significant decrease in the inexact channel matrix performance. Similarly, for the SPC recovery, the  ADMM network model (Algorithm \ref{alg:admm}), using the peak-signal-to-noise-ratio (PSNR, the higher to better) and the structural similarity index measure (SSIM, the closer to one the better) for various values of $\sigma$. The results show a decrease of up to 14 [dB] in the PSNR for $\sigma=0.9$ with respect to the ideal case. To address this issue, we propose a new learning approach to train the unrolling network using the mismatched fidelity term  \eqref{eq: mismatched}. \vspace{-0.4cm}

\begin{figure}[!t]
    \centering
    \includegraphics[width=0.92\linewidth]{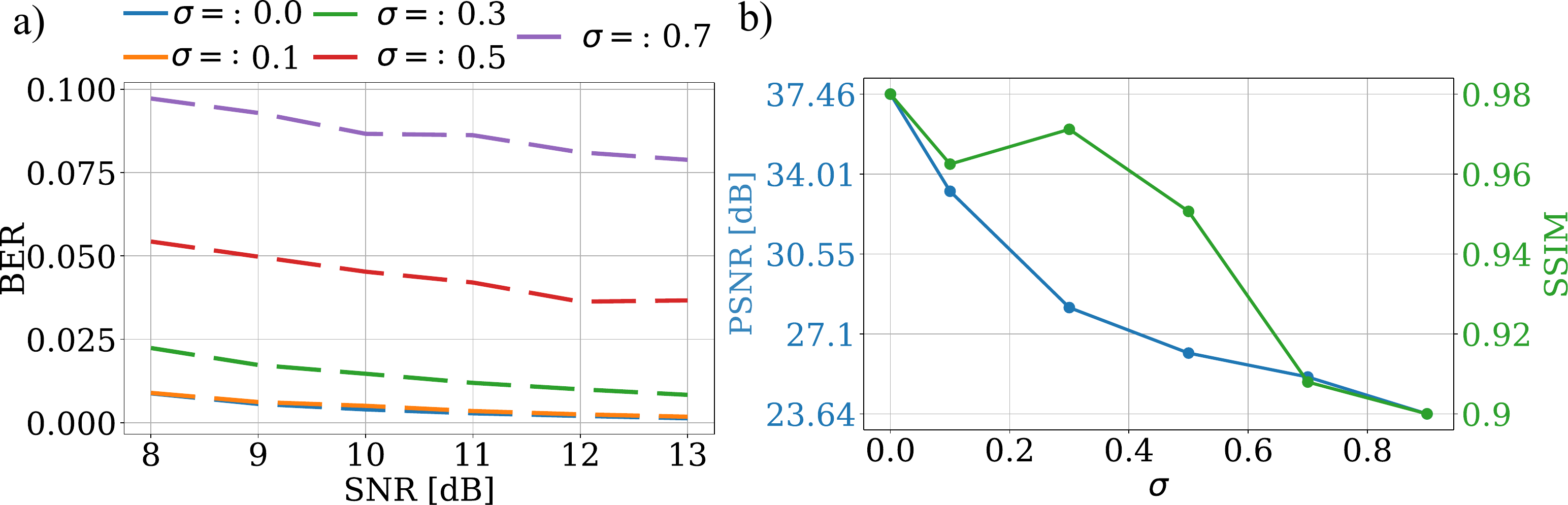}\vspace{-0.4cm}
    \caption{Effect of the variance in the unknown sensing operator $\mathbf{A}_u$ for a) deep MIMO detection with DetNet and b) SPC recovery with ADMM-Inspired network.}
    \label{fig:sigma}
\end{figure}

\subsection{Teacher structure}\vspace{-0.2cm}
Here we define the teacher as synthetic since we only have access to it in simulation.  Thus, let $\mathbf{A}_t = \mathbf{A}_{k} + \mathbf{A}_{ut}$ be the sensing matrix, where the entries of unknown operator $\mathbf{A}_{ut}$ {also follows} a Gaussian distribution $[\mathbf{A}_{ut}]_{i,j} \sim \mathcal{N}(0,\sigma_t^2)$ where $\sigma_t^2$ is the variance of the teacher, which satisfies that $0\leq \sigma_t^2< \sigma^2$. Note that the known matrix $\mathbf{A}_k$ is the same for the student and teacher system, the only difference relies on the variance of the unknown matrices. The corresponding measurements are  $\mathbf{y}_t = \mathbf{A}_t\mathbf{x}+\boldsymbol{\omega}$. The teacher model is trained based on the equation \eqref{eq:network_opt} with paired dataset $\{\mathbf{y}_t,\mathbf{x}\}$ leading to $\mathcal{M}_{\theta_t^*}(\mathbf{y}_t,\mathbf{A}_{k})$. Note, that in every iteration this network has access to the exact or almost-exact sensing operator since it has the structure of equation \eqref{eq:structure} which can lead to a better recovery performance (see Figure \ref{fig:sigma} in lower values of $\sigma$). In the proposed method, we can control the quality of the teacher by setting the parameter $\sigma_t$, thus this is a hyperparameter of the method. 
\vspace{-0.5cm}
\subsection{Distillation Loss Function}\vspace{-0.2cm}
To distill the knowledge of the teacher model to the student {model, we} propose to regularize the optimization of the student. Before introducing the proposed optimization problem, we introduce the following variables: the estimations at each stage of the student and teacher unrolling networks are denoted as $\tilde{\mathcal{X}}_s = \{\tilde{\mathbf{x}}_s^1,\dots,\tilde{\mathbf{x}}_s^L\}$ and $\tilde{\mathcal{X}}_s=\{\tilde{\mathbf{x}}_t^1,\dots,\tilde{\mathbf{x}}_t^L\}$. On the other hand, we define the collection of the data fidelity gradient at each stage for the student and teacher models as $\mathcal{G}_s = \{\nabla f(\tilde{\mathbf{x}}_s^1),\dots,\nabla f(\tilde{\mathbf{x}}_s^L)\}$ and  $\mathcal{G}_t = \{\nabla f(\tilde{\mathbf{x}}_t^1),\dots,\nabla f(\tilde{\mathbf{x}}_t^L)\}$. Thus, the proposed optimization is\vspace{-0.2cm} \small{
\begin{align}
    &\theta^* = \argmin_{\theta} \mathbb{E}_{\mathbf{x},\mathbf{y},\mathbf{y}_t} [\mathcal{L}(\mathcal{M}_\theta(\mathbf{y},\mathbf{A}_k),\mathbf{x},\mathcal{M}_{\theta^*_t}(\mathbf{y}_t,\mathbf{A}_{kt})]\\ &  
   \mathcal{L} ={\ell(\mathcal{M}_\theta(\mathbf{y},\mathbf{A}_k),\mathbf{x})} +{\ell_{KD}(\mathcal{M}_\theta(\mathbf{y},\mathbf{A}_k),\mathcal{M}_{\theta^*_t}(\mathbf{y}_t,\mathbf{A}_{k})})]\nonumber,\label{eq:network_opt}
\end{align}
}where the distillation loss is composed of two terms, one comparing the gradients of the student and teacher models and the other comparing their outputs. Then, distillation loss is
\begin{equation}
\ell_{KD} =  \lambda_{grad}{\ell_{grad} (\mathcal{G}_s,\mathcal{G}_t)}+ \lambda_{o}\ell_o(\tilde{\mathcal{X}}_s,\tilde{\mathcal{X}}_s),\vspace{-0.2cm}
\end{equation}
where $\lambda_{grad}$ and $\lambda_{o}$ are hyperparameters. Both loss functions are defined as\vspace{-0.5cm}
\begin{align}
\ell_{grad}(\mathcal{G}_s,\mathcal{G}_t))&=\sum_{i=1}^{L} \log(i)\Vert \nabla f(\tilde{\mathbf{x}}_s^i)-\nabla f(\tilde{\mathbf{x}}_t^i)\Vert_2,\\
\ell_{o}(\tilde{\mathcal{X}}_s,\tilde{\mathcal{X}}_t))&=\sum_{i=1}^{L} \log(i)\Vert\tilde{\mathbf{x}}_s^i- \tilde{\mathbf{x}}_t^i\Vert_2.
\end{align}
The main idea behind these loss functions is to guide the student's data fidelity gradient since the inexact sensing operator significantly affects this term. The output loss function guides the recovery at every iteration. An overview of the proposed method is shown in Fig. \ref{fig:pipeline}. {This approach involves higher computational costs during the student training phase, as it requires storing, performing inference, and pre-training the teacher model. However, only the student model is utilized during inference, ensuring that real-world deployment does not incur additional computational overhead.}
\begin{figure}[!t]
    \centering
    \includegraphics[width=0.85\linewidth]{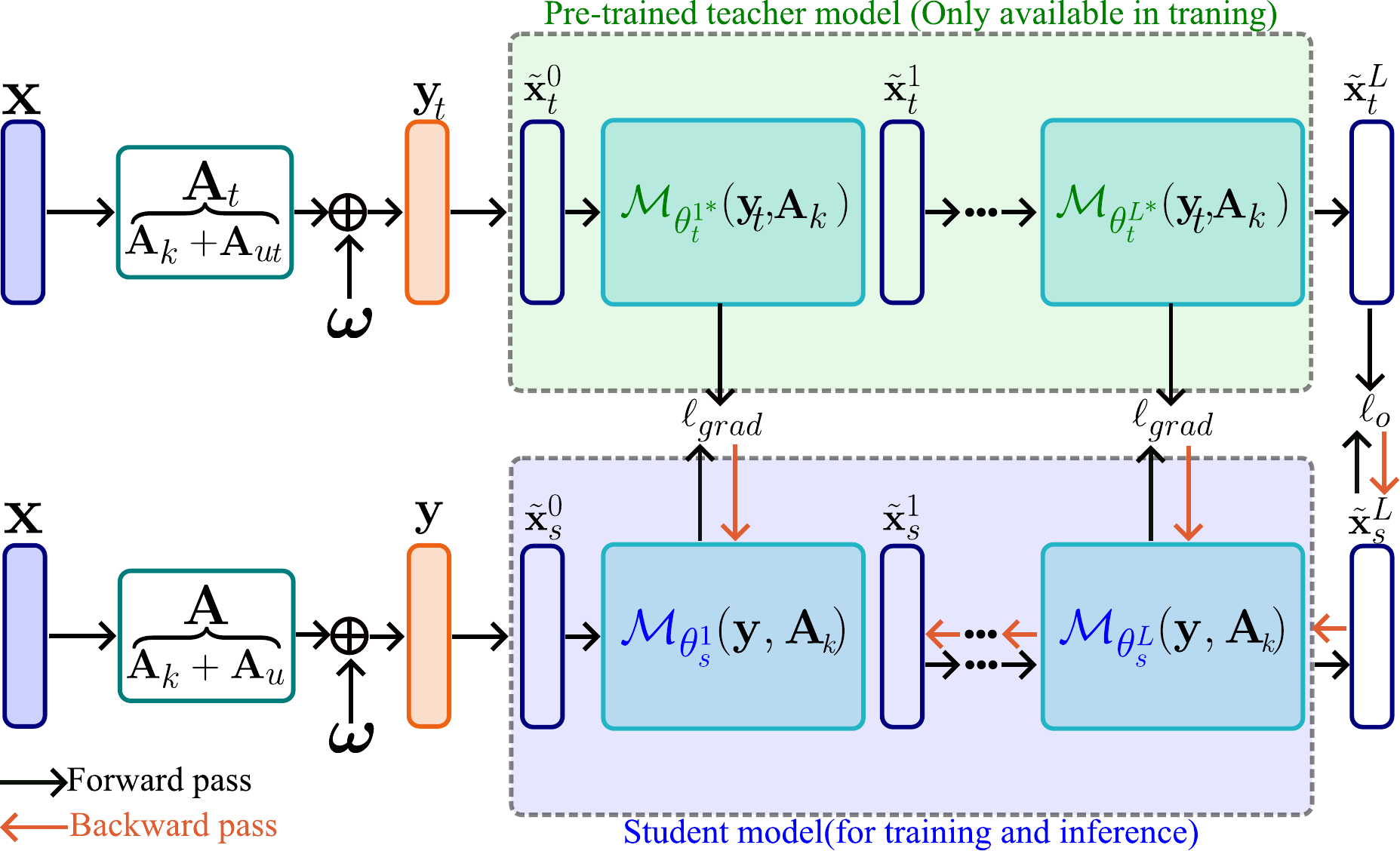}\vspace{-0.4cm}
    \caption{Scheme of the proposed optimization paradigm based on KD. The teacher model (green) consists of an unrolling network pre-trained with full or almost full knowledge of the sensing operator. In contrast, the student model (blue) consists of an unrolling network with the same architecture as the teacher's. However, the student only has access to partial knowledge of the sensing operator.} \vspace{0.2cm}
    \label{fig:pipeline}
\end{figure}

\section{Experiments \& Results}\vspace{-0.35
cm}
The effectiveness of the proposed approach was validated through several experiments in two key applications: SPC recovery and wireless communication via MIMO detection. The source code can be found in \cite{github}. The source code was developed in PyTorch framework \cite{paszke2019pytorch}. In all the results, the baseline is the student trained without the distillation loss functions. 

\textbf{SPC recovery:}
The MNIST dataset, consisting of 60,000 training images resized to $32 \times 32$ across, was used for training. The dataset was divided into $50,000$ images for training and $10,000$ for validation. The test set contains $10,000$ images. The unrolling network comprises $L=10$ stages and was trained for {$50$ epochs} using the Adam optimizer \cite{Adam} with a learning rate of  $5e^{-4}$ and a batch size of $600$ images. 
In this experiment, the known binary sensing matrix rows are the rows of the Hadamard basis using the cake-cutting ordering. The matrix was generated using the algorithm proposed in \cite{monroy2024hadamardrowwisegenerationalgorithm}. The parameters of the distillation loss functions were set to $\lambda_{grad} = \lambda_o = 1e^{-3}$. We used the mean-squared error (MSE) for the reconstruction loss function. We consider training several teacher models with different $\sigma_t = \{0.0, 0.1, 0.5, 0.7\}$ to validate our proposed method in this application.  Then, to train the student model, we consider $\sigma=\{0.3,0.5,0.7, 0.9\}$ and perform the training for each teacher model that satisfies that $\sigma>\sigma_t$. The result of this experiment is shown in Fig \ref{fig:spc}. Particularly, in Fig \ref{fig:spc}a) quantitative results are shown with a recovery performance mean of 10 repetitions, wherein each repetition new random unknown matrices $\mathbf{A}_u$ and $\mathbf{A}_{ut}$ are generated. The results show that the proposed KD approach improves the recovery PSNR by up to $1.5$[dB] in $\sigma = 0.3$ and $0.5$ [dB] in $\sigma=0.9$. Another observation is that the best performance for each $\sigma$ is not always the best teacher (the model with smaller $\sigma_t$) suggesting that we require a teacher with not too high performance to guide the student's training properly. In Fig \ref{fig:spc}b) is shown visual results of the proposed method and the baseline where the proposed method achieves better reconstruction quality for each value of $\sigma$.

\textbf{MIMO detection:}
 The DetNet was trained for 1,000 iterations. We set the number of transmitters to $n=30$ and the number of received antennas to $m=60$. The number of stages was set $L=90$. {In each iteration, random binary BPSK signals were generated along with the channel matrix which follows a Gaussian distribution. A batch size of 5,000 signals was used, and the additive Gaussian noise $\boldsymbol{\omega}$ with an SNR value sampled from a uniform distribution between 7 db and 13 db. Here the distillation losses hyperparameters were set to $\lambda_o =\lambda_{grad} = 1e^{-2}$. The recovery loss is the one used in the original DetNet paper \cite{samuel2017deep}. A linear-decay learning rate schedule was used, starting from $0.9e^{-3}$ with a decay factor of $0.97$ every $50$ iterations. 
 As in the first experiment, we study the effect of the values of $\sigma$ and $\sigma_t$ in the proposed method and compare them with the baseline. The results are shown in \ref{fig:mimo} where for different values of SNR in testing is displayed the detection BER of every configuration. The proposed method outperforms the baselines in the most challenging scenarios, with values of $\sigma\leq 0.5$. For $\sigma=0.3$ the baseline is better, however, this is because for this small value of $\sigma$ the detection performance is not significantly affected compared with the exact sensing operator model ($\sigma=0$ in Fig. \ref{fig:sigma}a)). Note that the best teacher setting for the student is not always the best-performing student. \vspace{-0.1cm}
\begin{figure}[!t]
    \centering
    \includegraphics[width=0.85\linewidth]{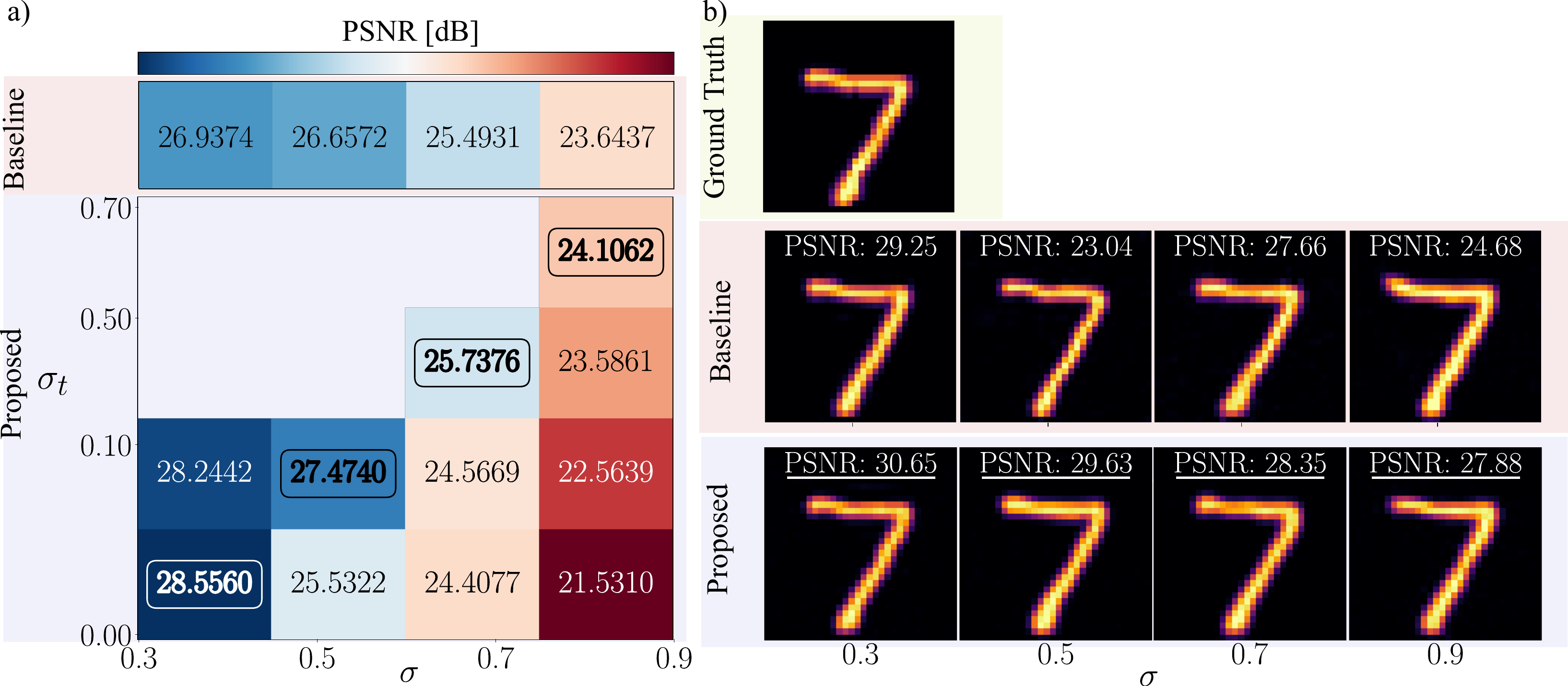}\vspace{-0.5cm}
    \caption{Recovery performance in the SPC system. a) quantitative results for different values of $\sigma_t$ and $\sigma$. b) Visual reconstruction for the best-performing configuration of the method and the baseline}\vspace{-0.05cm}
    \label{fig:spc}
\end{figure}

\begin{figure}[!t]
    \centering
    \includegraphics[width=0.85\linewidth]{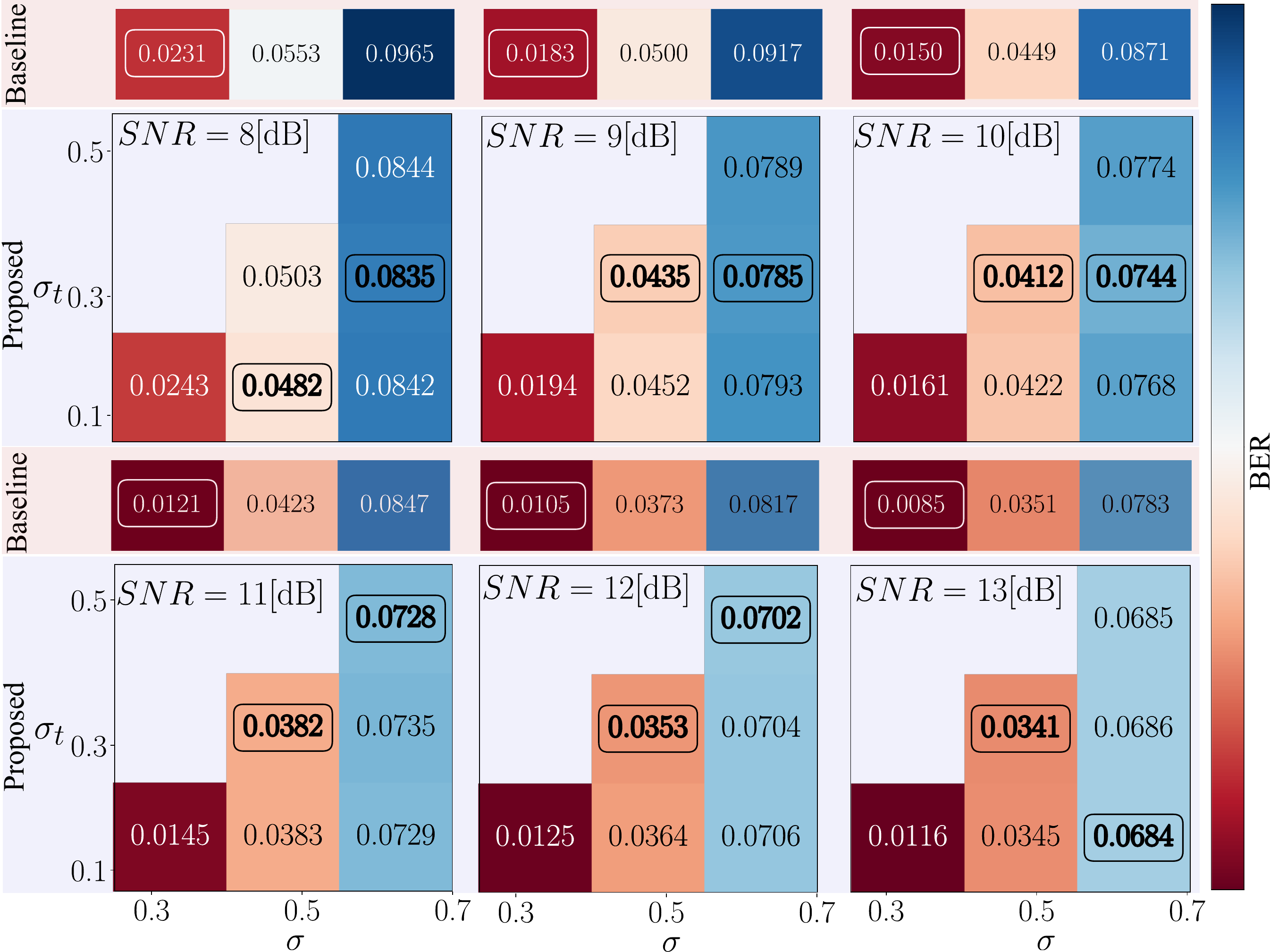}\vspace{-0.4cm}
    \caption{Detection performance in MIMO systems for different values of for the proposed method $\sigma_t$ and $\sigma$ and varying the SNR .} \vspace{0.2cm}
    \label{fig:mimo}
\end{figure}
\section{Conclusions}\vspace{-0.2cm}
A new recovery paradigm based on KD was proposed to address signal recovery methods with inexact sensing operators. This method leverages a system with a synthetic fully or nearly fully characterized sensing operator as the teacher model, which achieves high signal reconstruction performance. The teacher then transfers his knowledge to a system with a less characterized sensing operator, the student, using two distillation loss functions. Experimental results in SPC and MIMO detection validate the method's effectiveness. An important aspect of the proposed method is the teacher setting, where a high-performing teacher does not necessarily deliver a good knowledge transfer to the student. Future works will be focused on the teacher's design for optimal knowledge transfer. {Furthermore, this approach can be incorporated into models that address signal reconstruction with inexact sensing~operators~\cite{pu2021robust,qian2024robust} to enhance the quality and robustness of the reconstruction.}
}

\clearpage
\bibliographystyle{IEEEtran}
\footnotesize{\bibliography{biblio.bib}}

\end{document}